\documentclass[aps,pra,reprint,american,groupedaddress,amsmath,amssymb,showpacs,preprintnumbers,floatfix]{revtex4-1}

\usepackage[T1]{fontenc}
\usepackage[latin9]{inputenc}
\usepackage[letterpaper]{geometry}
\usepackage{array}
\usepackage{graphicx}
\usepackage{babel}
\usepackage{nicefrac}
\usepackage{hyperref}

\makeatletter
\newcommand{\sinc}{\mathrm{sinc}}
\newcommand{\avg}[1]{\langle #1\rangle}
\newcommand{\ket}[1]{| #1\rangle}
\newcommand{\bra}[1]{\langle #1|}
\newcommand{\vecb}[1]{\mathbf{#1}}
\makeatother

\begin{document}

\title{Eigenmode description of Raman scattering in atomic vapors in the presence of decoherence}

\author{Jan Ko{\l}ody\'nski, Jan Chwede\'nczuk, and Wojciech Wasilewski}

\affiliation{Faculty of Physics, University of Warsaw, ul. Ho\.{z}a 69, PL-00-681 Warszawa, Poland}

\begin{abstract}
A theoretical model describing the Raman scattering process in atomic vapors is constructed. The treatment investigates the low-excitation regime suitable for modern experimental applications. Despite the incorporated decoherence effects (possibly mode dependent) it allows for a direct separation of the time evolution from the spatial degrees of freedom. The impact of noise on the temporal properties of the process is examined. The model is applied in two experimentally relevant situations of ultra-cold and room-temperature atoms. The spatial eigenmodes of the Stokes photons and their coupling to atomic excitations are computed.  Similarly, dynamics and the waveform of the collective atomic state are derived for quantum memory implementations.
\end{abstract}


\maketitle

\section{Introduction}

Collective Raman scattering in multi-atomic ensembles serves as a potential, basic building block for quantum information processing and quantum state engineering \cite{Hammerer2010}. It can be used to generate single spin waves and photons \cite{VanderWal2003,Kuzmich2003}, store them \cite{Chaneliere2005} and shape their wavepackets \cite{Eisaman2004}. Recently, a major improvement in the control of the atomic ensembles has been achieved by means of multiple longitudinal spin wave modes, independent light wavepackets can be stored, reshuffled and reshaped \cite{Hosseini2009,Hosseini2011} or interfered with subsequent light pulses \cite{Campbell2012}. Moreover, transverse degrees of freedom can be used to generate spatially multimode squeezed states \cite{Boyer2008}. In this article, we formulate a theoretical model of the three-dimensional Stokes Raman scattering in atomic ensembles both in the ultra-cold and the room-temperature regimes. Despite the apparent complexity of the problem, we provide a simple method of obtaining the independently evolving eigenmodes, which fully determine the system's dynamics.

\begin{figure}[t]
  \includegraphics[clip, scale=0.25]{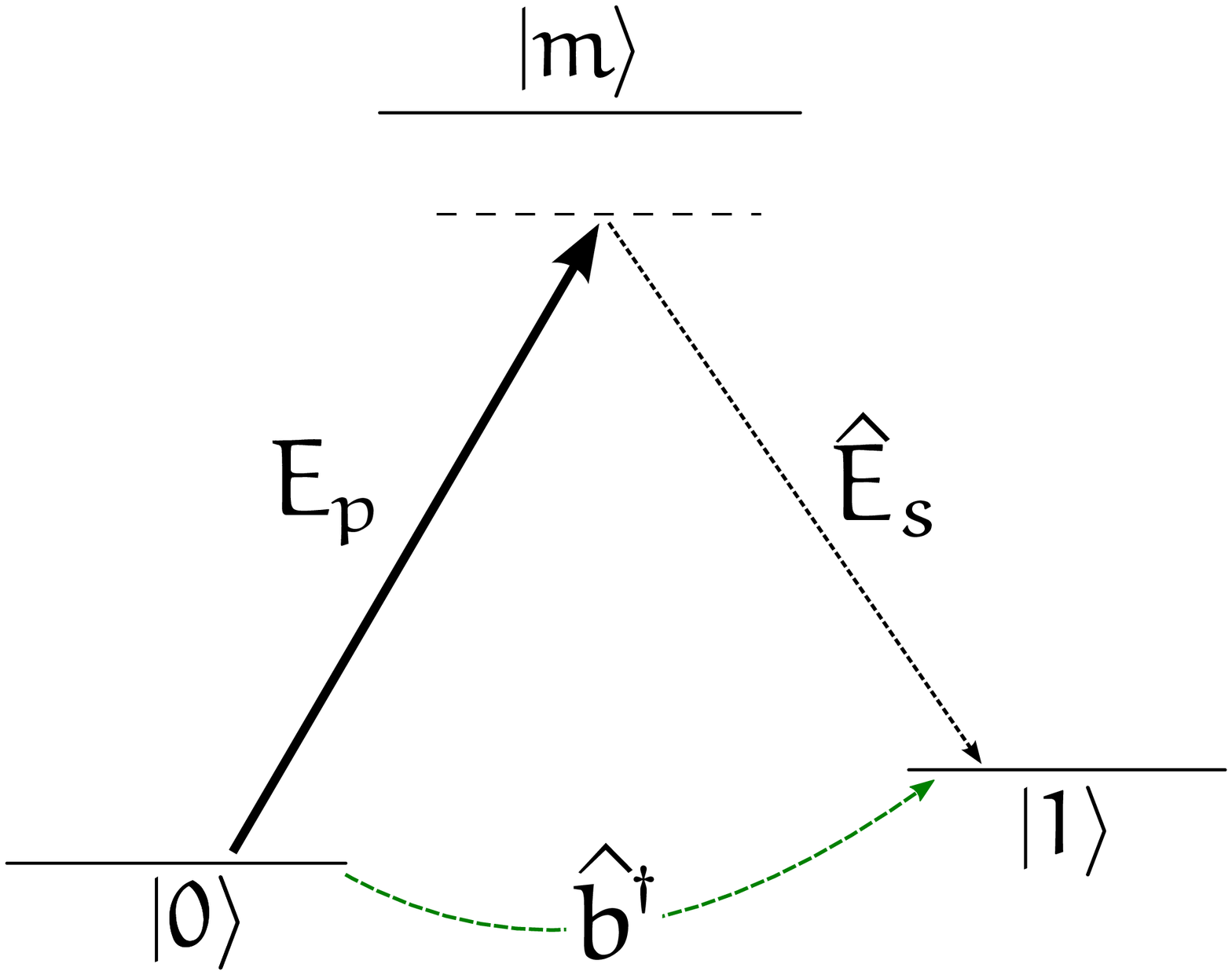}
  \caption{ (Color online)
Scheme of a Raman transition. An atom in state $\ket0$ is illuminated with an external pump field $E_p$, detuned from the $\ket0\rightarrow\ket m$ transition. 
    A single collective excitation, quantified by the operator $\hat b^\dagger$, is created by the $\ket0\rightarrow\ket1$ atomic transition. It is achieved owing to the absorption of a photon from the pumping beam and emission of a Stokes photon described by the field operator $\hat E_s$.}\label{scheme}
\end{figure}

Typically, the Raman scattering process occurs in an ensemble of atoms optically pumped into one of the ground states, say $\ket0$. Then, the collective excitations to another ground state, $\ket1$, called ``\emph{spin waves}'', can be utilized to store the quantum information. These excitations $\ket0\rightarrow\ket1$ are induced by the off-resonant coupling of $\ket 0$ to higher lying states, denoted by $\ket m$ in Fig.~\ref{scheme}, accompanied by the spontaneous emission of Stokes photons. The strength of the coupling is controlled by the Raman pump laser field, $E_p$. Furthermore, the read-out process can be easily performed by means of the inverse Raman scattering, in which the information stored by the excitations is passed onto the anti-Stokes photons produced \cite{DeEchaniz2008}. Here, we investigate  the dynamics of the information write-down alone. Nevertheless, our model can be easily reformulated to describe the read-out case.

Our treatment is complementary to the description introduced in the pioneering works \cite{Raymer1981,Mostowski1984,Raymer1985}, in which the macroscopic excitation regime was investigated. In contrast, we provide an exact solution in the low atomic excitation number limit, especially important for modern applications. Notably, we address the typical Gaussian transverse pumping beams, as opposed to the flattop profiles considered in \cite{Raymer1981,Mostowski1984,Raymer1985}. Furthermore, rather than focusing on the near field, we directly establish the far-field properties of scattered light, measurable in current experiments. In our approach, we neglect higher-order self-interaction effects, which lead to the build-up of the excitation gradient along the sample \cite{Raymer1985,Wasilewski2006}. Thanks to this simplification, which is valid in the regime of a low number of atomic excitations, we are able to decouple the problem's spatial and temporal degrees of freedom. This enables an in-depth analysis of the multimode capacity of the process. Hence, we provide the estimates of its eigenmode parameters, describe their evolution and include the possibly mode-dependent decoherence effects. In particular, we can easily account for the fact that diffusion most efficiently destroys the spin waves characterized by the short-distance spatial phase and amplitude variations. Crucially, we describe the final atomic state, which, serving as the quantum memory, could potentially be mapped onto the light in a read-out process \cite{Porras2008,Gorshkov2007a}.

This article is organized as follows. In Sec. \ref{model} we develop a general three dimensional model to obtain an eigenmode description of the system. In the eigenbasis, the Heisenberg equations coupling the atomic excitations with Stokes photons describe the pairwise coupling of the atomic and field modes, preserving the independence between distinct mode pairs. Moreover, this approach allows for the simple incorporation of the decoherence effects, typically diffusion and spontaneous emission. The Raman gain for each spin wave opposes the decoherence; thus by varying the pump intensity and the size of the sample one can change the number of spin-wave modes that are macroscopically occupied. 

Section \ref{temp_prop} discusses the temporal evolution of the statistics of each eigenmode. We show that, even for the low-occupied spin wave modes, the number of atomic excitations can be estimated with a precision that overcomes the quantum projection noise by counting the scattered photons. Hence, these results may facilitate experiments, in which the transverse multimode capacity of the scattering \cite{Boyer2008} would be combined with the storage \cite{VanderWal2003}.

Finally, in Sec. \ref{section_h}, we perform the spatial \emph{singular value decomposition} that is valid in two typical experimental situations (i.e. when an ultra-cold atomic cloud is interacting with a uniform Raman pump or when a room-temperature atomic cell is illuminated with a Gaussian pumping beam).

\section{The model \label{model}}

We consider an ensemble of identical atoms, each being initially in its ground state $\ket0$. Owing to the interaction of the probe with an intense, non-resonant, linearly polarized Raman pump beam with amplitude $\mathcal{E}_p$, frequency $\omega_p$ and wave vector ${\vecb k}_p$, that is,
\begin{equation}
  E_p(\vecb r,t)=\mathcal{E}_p(\vecb r,t)e^{i(\vecb k_p\cdot\vecb r-\omega_pt)}+\mathrm{c.c},
\end{equation}
some atoms are transferred to the other ground state $\ket1$. Each atomic transition is accompanied by an emission of a Stokes photon, as schematically depicted in Fig.~\ref{scheme}.

To analyze this process in detail, we describe the quantum state of Stokes photons and atoms respectively using two field operators
\begin{subequations}
\begin{eqnarray}
    \hat E^{(+)}_s(\vecb r,t)&=&\int\!\! d\vecb k\, e^{i(\vecb k\cdot \vecb r-ckt)} \hat a(\vecb k,t)\\
    \hat b(\vecb k,t)&=&\frac1{\sqrt N}\sum_\alpha e^{i(\vecb k \cdot \vecb r_\alpha-\omega_{01}t)} \ket0_\alpha\bra1_\alpha,\label{def_b}
\end{eqnarray}
\end{subequations}
where $c$ is the speed of light, $\omega_{01}$ is the $\ket0\rightarrow\ket1$ transition frequency and $k=|\vecb k|$. The operator $\hat b$ annihilates a quantum of the collective atomic excitation (i.e. a spin wave with momentum $\hbar\vecb k$). The index $\alpha$ runs over all the $N$ atoms in the ensemble.  If almost every atom occupies the $\ket 0$ state, the Holstein-Primakoff approximation \cite{Holstein1940} holds, so $\hat b$ satisfies the bosonic commutation relations.

When $N$ is large, one can replace the summation over separate particles in Eq.~(\ref{def_b}) with an integral over the atomic density $n(\vecb r)$. Within this approximation, the Hamiltonian of the Raman scattering process reads
\begin{equation}\label{ham}
  \hat H=i\hbar\iint\!\! d\vecb k\,d\vecb K\,h(\vecb k,\vecb K)\,\hat a^\dagger(\vecb k,t)\,\hat b^\dagger(\vecb K,t)e^{i\omega_k t}+\mathrm{H.c.},
\end{equation}
with $\omega_k=ck-\omega_s$, where we have introduced the central Stokes frequency $\omega_s=\omega_p-\omega_{01}$. The function $h(\vecb k,\vecb K)$ determines the probability amplitude of the photon-spin wave pair creation with wave vectors $\vecb k$ and $\vecb K$, respectively, and reads
\begin{equation}
  h(\vecb k,\vecb K)=\frac{g_0}{\hbar(2\pi)^3}\int\!d\vecb r\,e^{-i(\vecb k+\vecb K-\vecb k_p)\cdot \vecb r}\sqrt{n(\vecb r)}\,\mathcal{E}_p(\vecb r),\label{hfun}
\end{equation} 
where $g_0$ is the coupling constant 
\footnote{The coupling constant can be derived with the help of the adiabatic elimination of the excited states ${\ket m }$, \cite{Raymer1985}:
\begin{equation*}
    g_0=\sqrt{\frac{\omega_s}{\hbar\,\varepsilon_0}}\sum_md_{0m}d_{m1}\left(\frac1{\omega_{m0}+\omega_s}+\frac1{\omega_{m0}-\omega_p}\right), 
\end{equation*}
where ${d_{ij}}$ is the dipole moment of the ${|i\rangle\rightarrow|j\rangle}$ transition and ${\varepsilon_0}$ is the electric constant.}.
Note that the pump amplitude $\mathcal{E}_p$ has been chosen to be time independent, which corresponds to a common physical situation of square pulses. As this assumption is made only to simplify the calculations, one could in principle generalize the model for cases, when the pump amplitude $\mathcal{E}_p(\vecb r,t)$ is a separable function of $\vecb r$ and $t$.

The Hamiltonian (\ref{ham}), together with the commutation relations, gives the set of coupled Heisenberg equations, describing an evolution of the multimode fields of photons and atomic spin waves,
\begin{subequations}\label{general}
  \begin{eqnarray}
    &&\partial_t\hat a({\vecb k},t)=\int\! d\vecb K\,e^{i\omega_k t} h(\vecb k,\vecb K)\,\hat b^\dagger({\vecb K},t)\label{gena}\\
    &&\partial_t\hat b({\vecb K},t)=\int\! d\vecb k\,e^{i\omega_k t} h(\vecb k,\vecb K)\,\hat a^\dagger({\vecb k},t).\label{genb}
  \end{eqnarray}
\end{subequations}

\subsection{Modal decomposition} 

The modal decomposition of the fields is achieved by an adequate change of the basis for photons and spin waves. To demonstrate this, we start by integrating Eqs.~(\ref{gena}) and (\ref{genb}) up to the first order in the coupling constant $g_0$. The integral of $\exp(i\omega_k t)h(\vecb k,\vecb K)$ over an interval $t\!\in\![0,\tau]$ is proportional to $\sinc\!\left(\frac{\omega_k\tau}2\right)h(\vecb k,\vecb K)$. Let us note, that the sole $h(\vecb k,\vecb K)$ corresponds to an infinitely short interaction time ($\tau\!\rightarrow\!0$) and, in principle, would give rise to scattered photons of arbitrary frequency. Moreover, as indicated by Eq.~(\ref{hfun}), the width $w_k$ of $h(\vecb k,\vecb K)$ as a function of $\vecb k$ is inversely proportional to the effective spatial spread of the product $\sqrt{n(\vecb r)}\,\mathcal{E}_p(\vecb r)$. Hence, as in typical experiments the atomic sample is less than $10$-cm long, $w_k$ is constrained to be greater than $10\,\mathrm{m}^{-1}$. On the other hand, the $\sinc\!\left(\frac{\omega_k\tau}2\right)$ part is narrowly peaked around $k_s\!\equiv\!\omega_s/c$, so that even for an extremely short interaction time of $\tau\!\sim\!1\,\mathrm{ns}$ its width $(c\tau)^{-1}\ll w_k$. Thus, for any real interaction time, it is the $\sinc$ function that defines a thin spherical shell of allowed $\vecb k$ vectors for Stokes photons with radius $k\approx k_s$. Therefore, in spherical coordinates, the radius of the first argument of $h$ can be fixed at $k_s$. In other words, since the interaction lasts longer than the time it takes for photons to travel through the sample, the spectral width of the scattered light is independent of the shape of the atomic cloud. This is a significant step, which allows one to perform the singular value decomposition \cite{Golub1970} as follows:
\begin{eqnarray}
  &&\sinc\!\left(\frac{\omega_k\tau}2\right) h(\vecb k\equiv(k\!\approx\!k_s,k_\varphi,k_\theta),\vecb K)\;\;\;\;\;\;\;\;\;\;\nonumber\\
  &&=\frac{c\,\tau}{\pi k_s^2}\,\, \sinc\!\left(\frac{\omega_k\tau}2\right)\sum_{lm}\zeta_{lm}\,\psi^{\rm ph}_{lm}(k_\varphi,k_\theta)\,\psi^{\rm at}_{lm}(\vecb K) \label{svd}
\end{eqnarray}
with the $k$ dependence remaining only in the spectral factor. The integer numbers $(lm)$ parametrize the distinct modes and pair the mode functions of photons, $\psi^{\rm ph}_{lm}(k_\varphi,k_\theta)$, and spin waves, $\psi^{\rm at}_{lm}(\vecb K)$, which both form separate orthonormal bases. The real and positive singular values, denoted by $\zeta_{lm}$, quantify the coupling strength between  $\psi^{\rm ph}_{lm}$ and $\psi^{\rm at}_{lm}$ for each $(lm)$. The decomposition presented in Eq.~(\ref{svd}) is always possible for any regular coupling function and does not depend on the time interval $\tau$ provided it is sufficiently long.

\subsection{Uncoupled mode basis}

Using the mode functions from Eq.~(\ref{svd}), we expand the Stokes photon and the atomic spin wave annihilation operators respectively as
\begin{subequations}\label{decomp}
  \begin{eqnarray}
    \hat a(\vecb k,t)&=&\sum_{lm}\psi^{\rm ph}_{lm}(k_\varphi,k_\theta)\,\hat a_{lm}(k,t)\label{decompa} \\
    \hat b(\vecb K,t)&=&\sum_{lm}\psi^{\rm at}_{lm}(\vecb K)\,\hat b_{lm}(t),\label{decompb} 
  \end{eqnarray}
\end{subequations}
where we implicitly assume that the length of the photon $\vecb k$ vector varies only slightly around $k_s$. These expressions substituted into Eqs.~(\ref{gena}) and (\ref{genb}) give a pair of coupled equations for each independent mode $(lm)$, that is,
\begin{subequations}\label{dablm_dt}
  \begin{eqnarray}
    &&\partial_t\hat a_{lm}(k,t)=\zeta_{lm}e^{i\omega_k t}\,\hat b^\dagger_{lm}(t),\label{dalm_dt}\\
    &&\partial_t\hat b_{lm}(t)=\zeta_{lm}\!\!\int\!\! dk\, e^{-i\omega_k t}\,\hat a^\dagger_{lm}(k,t).\label{dblm_dt}
  \end{eqnarray}
\end{subequations}
Note that by taking the Fourier transform of Eq.~(\ref{dalm_dt}) in $k$, it becomes apparent that the interaction of the spatially extended Stokes field with the atomic excitations is modeled to be point-like. Furthermore, any higher order effects, such as the possibility of atomic excitations being carried over from one part of the sample into another, are excluded. In return, since the spin wave mode functions are time independent, we manage to separate the spatial and temporal degrees of freedom. Precise numerical investigations in one dimension \cite{Wasilewski2006} show that our approach is justified, as long as only a few excitations are present in the sample.

\subsection{Input-output relations of a mode pair}

The Eqs.~\eqref{dalm_dt} and \eqref{dblm_dt} resemble the analytically solvable Wigner-Weisskopf model of spontaneous emission. Their solution in time can be derived employing the ``time slicing'' method explicitly explained in Appendix~\ref{t_slice}. By introducing the operators denoting the Stokes field entering and leaving the atomic sample at time $t$, $\hat a^{\rm in}_{lm}(t)$, and $\hat a^{\rm out}_{lm}(t)$, the $k$-dependence is fully dropped and the final result reads
\begin{subequations}\label{fin}
  \begin{eqnarray}
    \hat a^{\rm out}_{lm}(t)&=&\int_0^t\!\!d\tau\left[\delta(t-\tau)+\zeta_{lm}^2e^{\frac12\zeta_{lm}^2(t-\tau)}\right]\hat a^{\rm in}_{lm}(\tau)\nonumber\\
    &&+\;\zeta_{lm}\,e^{\frac12\zeta_{lm}^2t}\,\hat b_{lm}^\dagger(0)\label{fina}\\
    \hat b_{lm}(t)&=&\int_0^t\!\!d\tau\;\zeta_{lm}\,e^{\frac12\zeta_{lm}^2(t-\tau)} \left[\hat a^{\rm in}_{lm}(\tau)\right]^\dagger\nonumber\\
    &&+\;e^{\frac12\zeta_{lm}^2t}\,\hat b_{lm}(0).\label{finb}
  \end{eqnarray}
\end{subequations}
The above set of equations, which preserve bosonic commutation relations, describes the process of the coupled amplification of Stokes and the atomic fields. The photons and spin waves are created in pairs with a rate defined by the singular values $\zeta_{lm}$ of the coupling function $h$, see Eq.~(\ref{svd}). This statement is no longer true in the presence of losses, which impact on the system's dynamics we investigate in the following section.

\subsection{Atomic decoherence}

The decoherence, which affects the atomic ensemble, is normally described by a combination of damping, caused by the random thermal motion and spontaneous emission of atoms, and space-uniform (illumination independent) losses, which are, for example, a consequence of atomic collisions. Neither the random thermal motion nor the uniform losses mix the distinct $(lm)$ modes. The rate of spontaneous emission, however, which is a function of the pumping beam's intensity, might vary between various parts of the sample. This, in principle, may lead to a coherent deformation of an atomic mode (i.e. a coherent cross-talk between different spin waves). In our analysis, we assume such effect to be negligible. 

The decoherence process, which does not mix the modes, is modeled by adding an expression $-\Gamma_{lm} {\hat b}_{lm}(t)+\sqrt{2\Gamma_{lm}} {\hat f}_{lm}(t)$ to Eq.~(\ref{dblm_dt}). The first term represents an exponential decay at the rate $\Gamma_{lm}$, whereas the second part is the Langevin noise and ensures the conservation of the commutation relations of both the photonic and atomic operators, since $[\hat f_{lm}(t),\hat f^\dagger_{l'm'}(t')]=\delta(t-t')\delta_{ll'}\delta_{mm'}$. 

The updated Eqs.~(\ref{dalm_dt}) and (\ref{dblm_dt}) can be integrated, giving the input-output relations of the Raman scattering in the presence of atomic decoherence, that is,
\begin{subequations}\label{damp}
  \begin{eqnarray}
    \hat a^{\rm out}_{lm}(t)&=&\int_0^t\!\!d\tau\left[\delta(t-\tau)+\zeta_{lm}^2e^{\gamma_{lm}(t-\tau)}\right]\hat a^{\rm in}_{lm}(\tau)\label{dampa}\\
    &+&\!\zeta_{lm}e^{\gamma_{lm} t}\,\hat b_{lm}^\dagger\!(0)\!+\!\zeta_{lm}\sqrt{2\Gamma_{lm}}\!\int_0^t\!\!\!d\tau\, e^{\gamma_{lm} (t-\tau)}\hat f_{lm}^\dagger\!(\tau)\nonumber\\
    \hat b_{lm}(t)&=&\zeta_{lm}\!\!\int_0^t\!\!d\tau\,e^{\gamma_{lm}(t-\tau)}\left[\hat a^{\rm in}_{lm}(\tau)\right]^\dagger\label{dampb}\\
    &+&\!e^{\gamma_{lm}t}\,\hat b_{lm}(0)+\sqrt{2\Gamma_{lm}}\!\int_0^t\!\!d\tau\, e^{\gamma_{lm} (t-\tau)}\hat f_{lm}(\tau),\nonumber
  \end{eqnarray}
\end{subequations}
where $\gamma_{lm}={\zeta_{lm}^2}/2-\Gamma_{lm}$. 

We expect a threshold behavior: When $\zeta_{lm}^2>2\Gamma_{lm}$, the pair creation rate exceeds the losses, giving exponential growth in time of the population of the atom-spin wave pairs. In the next section, based on Eqs.~(\ref{dampa}) and (\ref{dampb}), we discuss in more detail both the temporal and spatial coherence of the system.

Finally, let us briefly comment on how our result can be extended to describe other types of Raman scattering. For example, if the pump's frequency was tuned to level $\ket1$ instead, the interaction would lead to an exchange of excitations between the light and atoms, as generated by an ${\hat a}^\dagger \hat b$ type Hamiltonian rather than the one of Eq.~(\ref{ham}). Thus, one can construct the correct input-output relations from our result in a straightforward manner by removing the daggers from all the operators in Eqs.~(\ref{dampa}) and (\ref{dampb}) and substituting $\zeta_{lm}\rightarrow i\zeta_{lm}$. Moreover, in the case of the pumping beam coupled to both $\ket0$ and $\ket1$, following a similar procedure, one would recover the equations described in \cite{WasilewskiOE2009}.

\section{Temporal properties of photons and atoms \label{temp_prop}}

Since the Hamiltonian (\ref{ham}) is quadratic and the initial state is a vacuum, then all the temporal properties of the system are fully determined by the following set of correlation functions:
\begin{subequations}\label{cor_fun}
  \begin{eqnarray}
    \avg{\hat a^\dagger\!(t)\, \hat a(t')}
    &=&\frac{\zeta^2}\gamma\left((\Gamma+\gamma)e^{\gamma(t+t')}-\Gamma e^{\gamma|t-t'|}\right)\;\;\;\;\;\;\;\;\;\label{cor_fun_a} \\
    \avg{\hat b^\dagger\!(t)\, \hat b(t')}&=&\frac{\zeta^2}{2\gamma}\left(e^{\gamma(t+t')}-e^{\gamma|t-t'|}\right)\label{cor_fun_b} \\
    \avg{\hat b(t)\,\hat a(t')}&=& \frac\zeta\gamma\left((\Gamma+\gamma)e^{\gamma(t+t')}-\Gamma e^{\gamma|t-t'|}\right).
  \end{eqnarray}
\end{subequations}
Because the separate modes of Eqs~(\ref{damp}) evolve independently, the mode indices $(lm)$ were dropped in the above equations without loss of generality. Also, as only the output photons are considered, we have introduced $\hat a(t)=\hat a^{\rm out}(t)$.

In the decoherence-free case, when $\Gamma=0$, the three correlation functions (\ref{cor_fun}) factorize. Hence, all the temporal properties are determined by a single mode $\propto~\!\!\exp[(\zeta^2 t)/2]$. Moreover, at any time $t$, the average cumulative number of scattered photons is equal to the mean number of atomic excitations present, that is, 
\begin{equation}
  \avg{\hat b^\dagger\!(t)\, \hat b(t)}=\int\limits_{0}^t\!\!d\tau\,\avg{\hat a^\dagger\!(\tau) \,\hat a(\tau)}=e^{\zeta^2 t}-1.\label{eq}
\end{equation}
This expression indicates that the system can be described as a sum of independent squeezed states for each $(lm)$ mode, since Stokes photons and spin waves scatter in correlated pairs. 
\begin{figure}[t]
    \includegraphics[width=1\columnwidth]{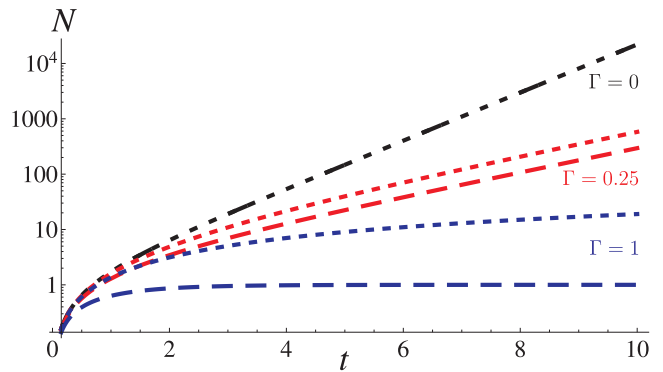}  
    \caption{(Color online) The mean number $N$ of scattered photons (\emph{dotted lines}) and spin waves (\emph{dashed lines}), as a function of $t$, for $\zeta=1$ and three different values of decoherence strength $\Gamma$. When $\Gamma=0$, the numbers of the atomic excitations and Stokes photons are equal (\emph{black dash-dotted line}). The curves flatten with increasing $\Gamma$ and for strong enough damping, here $\Gamma>1/2$, the number of spin waves saturates (\emph{blue dashed line}).
    }\label{one_mod}
\end{figure}

The situation changes dramatically in the presence of decoherence, when $\Gamma\neq0$. The correlation functions (\ref{cor_fun}) no longer factorize, thus the fields $\hat a(t)$ and $\hat b(t)$ become multimode. First, we investigate the impact of decoherence on the mean number of scattered atoms and photons for various values of $\Gamma$ with fixed $\zeta=1$, see Fig.~\ref{one_mod}. Clearly, with the strength of decoherence rising, the rate at which excitations are created drops. Furthermore, the mean number of scattered atoms behaves in two distinct manners. When $\Gamma\le\zeta^2/2$, then $\gamma\ge0$ and $\langle\hat b^\dagger\!(t)\,\hat b(t)\rangle$ grows exponentially in time. It saturates, on the other hand, for $\Gamma>\zeta^2/2$, which can be easily inferred from Eq.~(\ref{cor_fun_b}). On the contrary, the average number of Stokes photons obtained by integrating Eq.~(\ref{cor_fun_a}) increases irrespective of $\Gamma$. 

\begin{figure}[t]
    \includegraphics[width=1\columnwidth]{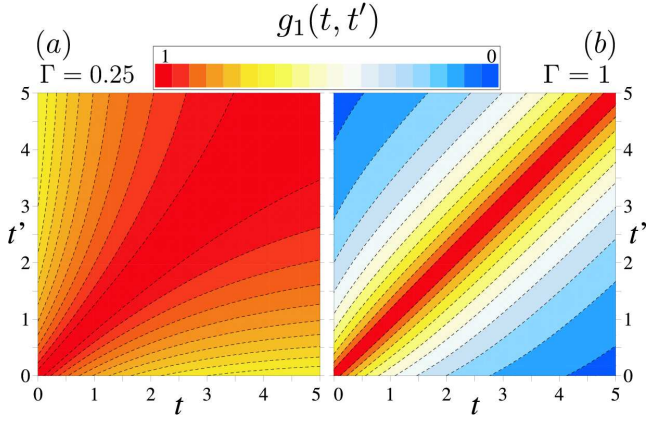} 
    \caption{(Color online) The normalized temporal correlation function of scattered photons, $g_1(t,t')$, for the subcritical and supercritical losses regimes [i.e., $\Gamma\le1/2$ and $\Gamma>1/2$, ($\zeta=1$)].
    }\label{temp_cor}
\end{figure}

Second, to examine the effect of decoherence on the system in more detail, we focus on the normalized temporal correlation function of photons, 
\begin{equation}
  g_1(t,t')=\frac{\avg{\hat a^\dagger\!(t)\, \hat a(t')}}{\sqrt{\avg{\hat a^\dagger\!(t)\, \hat a(t)}\avg{\hat a^\dagger\!(t')\, \hat a(t')}}}, \label{norm_ph}
\end{equation}
which can be directly accessed in an experiment. In the lossless case, only one temporal mode is present, hence for any $t$, $t'$: $g_1(t,t')=1$, which confirms the full time coherence of the system. When decoherence effects are included, the two distinct regimes are again observed. In the subcritical case of $\Gamma\le\zeta^2/2$, [e.g., Fig.~\ref{temp_cor}$(a)$], the losses are not large enough to completely suppress the gain. Therefore, $g_1$ always increases up to 1, as one moves upwards along the antidiagonal direction on the diagram. On the contrary, in the supercritical losses regime of $\Gamma>\zeta^2/2$ [e.g., Fig.~\ref{temp_cor}$(b)$], the full time coherence only persists on the antidiagonal [i.e., for all $t\!\ge\!0$: $g_1(t,t)=1$] and the correlation function always exponentially drops in the antidiagonal directions. Looking back at Eq.~(\ref{cor_fun_a}) with $\gamma<0$  and calculating $ \lim_{(t+t')\rightarrow\infty} g_1(t,t')=e^{\gamma|t-t'|}$, we realize that, in the supercritical regime, $g_1$ decays away from the antidiagonal even infinitely far from the origin $t=t'=0$.

\subsection{Estimation of the number of atomic excitations}

In a typical scattering experiment, the number of Stokes photons is measured as a function of time. Based on this information, one can estimate the number of atomic excitations created in the process. Here, we demonstrate how to maximize the precision of the estimation procedure by utilizing the correlation functions of Eq.~(\ref{cor_fun}).

%We have shown that, on average, in the ideal lossless case the number of atomic excitations is simply equal to the total number of Stokes photons detected, whereas in the presence of decoherence these two numbers differ, see Fig.~\ref{one_mod}.
To construct such an estimation strategy (optimal for any instance of an experiment of duration $T$) we need to minimize the mean squared difference between the probability distributions describing the real and the estimated numbers of spin waves created in the process. 
%Our goal is to construct a strategy of estimating $n_b=\langle\hat b^\dagger\!(T)\,\hat b(T)\rangle$ based on results of measurement of $\langle\hat a^\dagger\!(t)\,\hat a(t)\rangle$. 
To this end, we construct an operator representing the statistical distance between those distributions, that is,
\begin{equation}
 \hat{\mathcal D}=\hat b^\dagger\!(T)\,\hat b(T)-\int_0^T\!\!\!dt\;\hat a^\dagger\!(t)\,\hat a(t)\,w(t),
\end{equation}
where different estimation procedures correspond to various time-dependent weight functions  $w(t)$. Then, the optimal strategy is represented by $w(t)$, which minimizes the mean squared error, that is,
\begin{eqnarray}
	\avg{\hat{\mathcal D}^2} &=& n_b(n_b+2)
	+\int_0^T dt\,dt'\,w(t)w(t')\avg{\hat a^\dagger\!(t)\hat a(t')} \nonumber\\ 
	&+&\int_0^T dt \, w(t)[w(t)-2 n_b]\avg{\hat a^\dagger\!(t)\hat a(t)} \nonumber\\
	&-&2 \int_0^T dt \, |\avg{\hat b(T)\hat a(t)}|^2 w(t) \nonumber\\
	&+&\left(\int_0^T dt\, w(t) \avg{\hat a^\dagger\!(t)\hat a(t)}\right)^2
\end{eqnarray}
with $n_b\!=\!\avg{\hat b^\dagger\!(T)\,\hat b(T)}$ being the mean number of spin waves present at $T$. Hence, the optimal $w(t)$ should satisfy the differential criterion $\delta \avg{\hat{\mathcal D}^2}/\delta w=0$, which corresponds to the equation
\begin{eqnarray}\label{w_cond}
 	&&\int_0^T\!\!\!dt'\,\left( |\avg{\hat a^\dagger\!(t)\hat a(t')}|^2 +  \avg{\hat a^\dagger\!(t)\hat a(t)}\avg{\hat a^\dagger\!(t')\hat a(t')} \right)w(t') \nonumber\\
 	&&+w(t)\avg{\hat a^\dagger\!(t)\hat a(t)}=|\avg{\hat b(T)\hat a(t)}|^2+ n_b \avg{\hat a^\dagger\!(t)\hat a(t)}
\end{eqnarray}
that is fully determined by the correlations functions of Eq.~(\ref{cor_fun}).
\begin{figure}[t]
   \includegraphics[width=1\columnwidth]{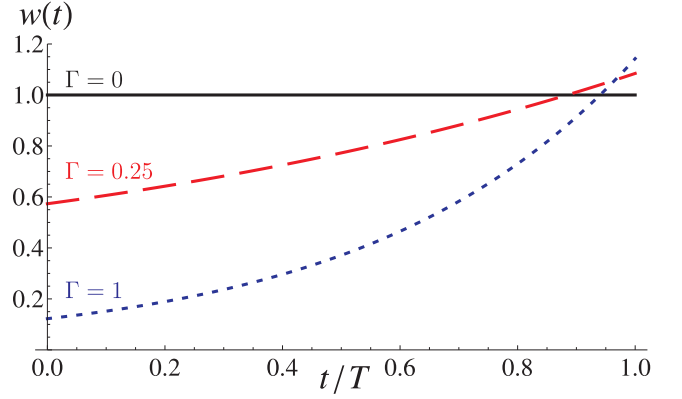}
   \caption{(Color online) The optimal weight function $w(t)$ for estimating the number of atomic excitations present at time $T$, when the pumping beam is switched off, ($\zeta\!=\!1$). In the lossless case of $\Gamma\!=\!0$, $w(t)\!=\!1$, as the average numbers of spin waves and detected Stokes photons coincide at all times and no estimation is required. The optimal weight function becomes steeper, as the strength of decoherence rises.
   }\label{opt_est}
\end{figure}

The above equation can be easily solved numerically by approximating all of the terms piecewise linearly. The result is a rising function, as illustrated in Fig.~\ref{opt_est}. Its shape can be explained by realizing that for a pump pulse of duration $T$ we expect some excitations produced at the beginning of the interaction $t\simeq 0$ to wash out, while those produced at $t\simeq T$ to generally prevail. Therefore, it is indeed judicious to observe the photon counts registered at the beginning less than those produced at the end of scattering.

It is interesting to note that the application of the proper weight function will lead to a sub-Poissonian (sub-projection-noise) precision of the estimation of the number of atomic excitations (i.e. $\avg{\hat{\mathcal D}^2}<n_b$) for any setting. For example, although for large losses $\zeta=1$, $\Gamma=1$, and very long pulse duration $T>10$ the mean number of atomic excitations and the average total number of scattered photons are only vaguely correlated, using a weight function $w(t)=1.24 \exp[2.23(t-T)]$ one can estimate the number of atomic excitations with noise over 3-dB smaller than the projection noise.

\section{Spatial properties of photons and atoms}\label{section_h}

To obtain a complete picture of the Raman scattering process, one needs to know the spatial functions $\psi^{\rm ph}_{lm}(k_\varphi,k_\theta)$ and $\psi^{\rm at}_{lm}(\vecb K)$ used in the original $(lm)$ mode decomposition of Eq.~(\ref{decomp}). These mode functions, of Stokes photons and spin waves respectively, were defined in Eq.~(\ref{svd}) by the \emph{singular value decomposition} of the coupling function $h(\vecb k,\vecb K)$. In this section we demonstrate how they can be easily found in typical experimental conditions. We focus on the two situations: The case of cold atoms, when the shape of atomic cloud is characterized with a Gaussian distribution, and the room-temperature case, when atoms are uniformly distributed in a cell. Although the final form of $\psi^{\rm ph}_{lm}$ and $\psi^{\rm at}_{lm}$ has to be numerically computed, its accuracy is assured by many steps that are analytically performed in the construction procedure.

\subsection{Coupling function $h(\vecb k,\vecb K)$}

The coupling function $h$ is defined in Eq.~(\ref{hfun}) via a Fourier transform of the product $\sqrt{n(\vecb r)}\,\mathcal{E}_p(\vecb r)$, where $n(\vecb r)$ stands for the atomic density and $\mathcal{E}_p(\vecb r)$ is the amplitude of the classical pump. Since the system has a rotation symmetry around the direction of the pumping beam, taken to be $z$, it is sensible to describe the process in cylindrical coordinates $\vecb{r}\equiv(\boldsymbol\rho\!\equiv\!(\rho,\varphi),z)$.

({\bf a}) First, for the case of atoms at \emph{low temperatures}, we assume their density to be distributed according to a Gaussian, that is,
\begin{equation}
  n(\vecb r)=n(\rho,z)=\left(\frac2\pi\right)^{\frac32}\frac{n\,V}{\sigma_{a,\rho}^2\,\sigma_{a,z}} e^{-\frac{2\rho^{2}}{\sigma_{a,\rho}^{2}}-\frac{2z^{2}}{\sigma^2_{a,z}}},
\end{equation}
where $V$ is the volume of the probe, $n V$ is the total number of atoms, and $\sigma_{a,\rho|z}$ effectively parametrize the spread of the atomic cloud. The amplitude of the pumping beam is taken then to be uniform, hence $\mathcal{E}_p(\vecb r)=\sqrt{I_0/\varepsilon_0}$ with $I_0$ being the pump's intensity.

({\bf b}) Second, in the \emph{room-temperature} case, we assume the atomic density to be constant and equal to $n$, but confined to some finite axial region $|z|\leq L/2$ with $L$ being the length of the atomic cell. We take the amplitude of the pumping beam to be Gaussian, hence
\begin{equation}
  \mathcal{E}_{p}(\vecb{\rho},z)=\sqrt{\frac{I_{0}}{\varepsilon_{0}}}e^{-\frac{\rho^2}{\sigma_p^2(1+\xi_p^2)}}e^{i\frac{\rho^2}{\sigma_p^2}\frac{\xi_p}{1+\xi_p^2}},
\end{equation}
where $\xi_p=2z/(k_p\sigma_p^2)$ and $\sigma_p$ is the beam's waist size. 

Both cases lead to the same general coupling function 
\begin{equation}\label{h_kK_gen}
  h(\vecb k,\vecb K)=\frac{g_0}{\hbar(2\pi)^3}\int\!\! d\vecb r\, e^{-i (\vecb k+\vecb K -\vecb k_p)\cdot\vecb r}e^{-\frac{\rho^2}{\sigma^2(1+i \xi)}} Z(z), 
\end{equation}
for which, to recover the regime ({\bf a}), one has to set: $\xi=0$ and $\sigma=\sigma_{a,\rho}$, whereas for ({\bf b}): $\xi=\xi_p$ and $\sigma=\sigma_{p}$. The function $Z(z)$ is then defined as follows
\begin{equation}
  Z(z)=\sqrt{\frac{nI_{0}}{\varepsilon_{0}}}\times
  \begin{cases}
    \sqrt{\left(\frac2\pi\right)^{\frac32}\frac{V}{\sigma_{a,\rho}^2\,\sigma_{a,z}}}e^{-\frac{z^2}{\sigma^2_{a,z}}} & \!\!\!\!\!\!\text{,\;\;\;for ({\bf a})} \\
    1 & \!\!\!\!\!\!\text{,\;\;\;for ({\bf b})}
  \end{cases}.
\end{equation}

In real experimental setups, one calibrates the detection system to focus on the Stokes photons that are most probably produced in the process but still can be efficiently measured. Normally, these are the ones produced along the atomic sample at small angles to the direction of the pumping beam. Hence, using again the fact that the length of the Stokes photon wave vector is approximately fixed at $k=|\vecb k|\approx k_s$, we apply the paraxial approximation as $k_z\approx k_s-k_\rho^2/(2k_s)$ in  the reference frame, in which the atomic wave vector transforms to ${\vecb K}\rightarrow{\vecb K}+{\vecb k}_p-k_s\vecb e_{z}$. Moreover, to diminish the number of free parameters, we introduce the dimensionless variables in the position space: 
$z=\sigma_z\tilde z$, $\boldsymbol{\rho}=\sqrt{\sigma_z/k_s}\,\tilde{\boldsymbol{\rho}}$ and similarly for the wave vectors $\boldsymbol{\kappa}\!=\!\{\vecb k,\vecb K\}$: $\boldsymbol{\kappa}_{\boldsymbol{\rho}}=\sqrt{k_s/\sigma_z}\,\tilde{\boldsymbol{\kappa}}_{\boldsymbol{\rho}}$, $\kappa_z=(1/\sigma_z)\,{\tilde \kappa}_z$, with $\sigma_z$ quantifying the effective size of the atomic ensemble in the $z$ direction.

Finally, we obtain the revised coupling function
\begin{eqnarray}\label{h_kK_dimless}
  h\!\left(\tilde{\vecb k},\tilde{\vecb K}\right)&=&\frac{g_0}{\hbar(2\pi)^3}
  \int \!\! d\tilde{\vecb r} \,
  e^{-i\left[ (\tilde{\vecb k}_{\boldsymbol{\rho}}+\tilde{\vecb K}_{\boldsymbol{\rho}})\cdot\tilde{\boldsymbol{\rho}}
    +(\tilde K_z-{\tilde k}_\rho^2/2)\tilde z\right]} \nonumber\\
  &\times& e^{-{\tilde\rho^2}/({2\pi F\![\sigma,\sigma_z]s + i2\kappa\tilde z})} \;\tilde Z(\tilde z),
\end{eqnarray}
where now, to retrieve the case ({\bf a}), one has to set: $\sigma=\sigma_{a,\rho}$, $\sigma_z=\sigma_{a,z}$ and $\kappa=0$, whereas for ({\bf b}): $\sigma=\sigma_{p}$, $\sigma_z=L$ and $\kappa=k_s/k_p\approx1$. The transformed $\tilde Z(\tilde z)$ function now reads
\begin{equation}
  \tilde Z(\tilde z)=\sqrt{\frac{nI_{0}}{\varepsilon_{0}k_s^2}}\times
  \begin{cases}
    \sqrt{\left(\frac2\pi\right)^{\frac32}\frac{V\sigma_{a,z}^3}{\sigma_{a,\rho}^2}}\;e^{-\tilde z^2} & \!\!\!\!\!\!\text{,\;\;\;for ({\bf a}),} \\
    L^2\text{,} & \!\!\!\!\!\!\text{\;\;\;for ({\bf b}).}
  \end{cases}
\end{equation}
Most importantly, Eq.~(\ref{h_kK_dimless}) up to a multiplicative factor depends only on a single parameter, the \emph{Fresnel number}, of the illuminated portion of the sample
\begin{equation}
  F=\frac{k_s \sigma^2}{2\pi \sigma_z}=\frac{k_s}{2\pi}\times
  \begin{cases}
      \sigma_{a,\rho}^2/\sigma_{a,z} & \!\!\!\!\!\text{,\;\;\;for ({\bf a}),} \\
      \sigma_p^2/L\text{,} & \!\!\!\!\!\text{\;\;\;for ({\bf b}),}
  \end{cases}
\end{equation}
which is fully determined by the geometry of the probe. 

In conclusion, having fixed the value of $F$ and performed the singular value decomposition of $h(\tilde {\vecb k}, \tilde{\vecb K})$ defined in Eq.~(\ref{h_kK_dimless}), we can simply establish the required modes of $h(\vecb k, \vecb K)$, see Eq.~(\ref{svd}), just by shifting back the $\vecb K$ vector and rescaling the obtained eigenmodes.

\subsection{Mode functions of photons and atoms}

We are now in a position to evaluate the spatial mode functions $\psi_{lm}^{\rm ph}$ and $\psi_{lm}^{\rm at}$ of the Stokes photons and atomic excitations used in Eq.~(\ref{decomp}). First, we express them in orthonormal functional bases. In that way, the singular value decomposition of the coupling function $h(\vecb k,\vecb K)$ can be numerically performed on a discrete tensor of the basis coefficients. As argued in Appendix \ref{decomph}, this tensor can be computed analytically in the position space using the Laguerre-Gaussian modes as the basis functions for both $\psi_{lm}^{\rm ph}$ and $\psi_{lm}^{\rm at}$. As the result of this procedure, we obtain the Fourier transform, $\mathcal{F}$, of Eq.~(\ref{svd}):
\begin{equation}\label{ft_svd}
  {\mathcal{F}}\!\left[\!\frac{c\,\tau}{\pi k_s^2}\,\sinc\!\left(\frac{\omega_k\tau}2\right) h(\vecb k,\vecb K)\right]\!=\!\sum_{lm}\!\zeta_{lm}\,\psi^{\rm ph}_{lm}(\vecb r)\,\psi^{\rm at}_{lm}(\vecb r'), 
\end{equation}
where the mode functions in the paraxial approximation, owing to the cylindrical symmetry of the left-hand side, can be written as
\begin{subequations}\label{mod_pos_exp}
  \begin{eqnarray}
    \psi^{\rm ph}_{lm}(\vecb r)&=&e^{ik_sz}\,e^{im\varphi}\,\psi^{\rm ph}_{lm}(\rho,z)\\
    \psi^{\rm at}_{lm}(\vecb r)&=&e^{-ik_sz}\,e^{i{\vecb k}_p\cdot\vecb r}\,e^{-im\varphi}\,\psi^{\rm at}_{lm}(\rho,z).
  \end{eqnarray}
\end{subequations}
Their relation to the original $\psi_{lm}^{\rm ph}(\vecb k)$ and $\psi_{lm}^{\rm at}(\vecb K)$ can be found in Appendix \ref{decomph}.
\begin{figure}[t]
 \includegraphics[width=1\columnwidth]{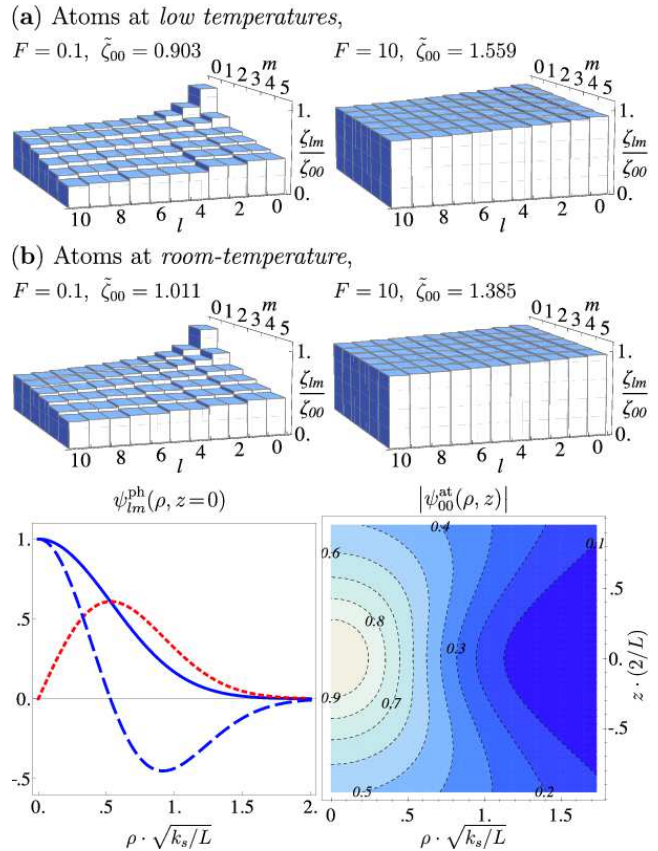}
 \caption{(Color online) The singular values $\zeta_{lm}$ of dominant $(lm)$ modes calculated for the Fresnel numbers $F\!=\!\{0.1,10\}$ in the cold atoms and the thermal atomic vapor regimes. In the latter case and $F\!=\!10$, we also plot the photonic eigenmodes $\psi^{\rm ph}_{lm}(\rho,z\!=\!0)$ for $m=0$, $l=0$ (solid blue line), $m=0$, $l=1$ (dashed blue line) and $m=1$, $l=0$ (dotted red line), and the modulus of the first atomic mode function, $\left|\psi_{00}^{\rm at}(\rho,z)\right|$.}\label{mod_funs}
\end{figure}

Figure~\ref{mod_funs} shows the dominant singular values $\zeta_{lm}$ for the Fresnel numbers $F=\{0.1,10\}$  in both ({\bf a}) and ({\bf b}) regimes. In the ultra-cold atoms case, ({\bf a}), we restrict ourselves in numerics to the axial range $|z|\le3\,\sigma_{a,z}$, so that the Gaussian distributed atomic cloud is well inside the sampled region. On the other hand, at room temperatures ({\bf b}), as we consider Stokes photons scattered nearly with the momentum of the pumping beam, we set $\kappa\!\equiv\!k_s/k_p=1$. Furthermore, in this case we plot the first three photonic and the first atomic eigenmode for $F=10$. 
In both regimes, we explicitly write the primary unnormalized singular values, ${\tilde \zeta}_{00}$, with numerical precision to the third decimal place, which are related to the previously introduced $\zeta_{00}$ via rescaling, that is,
\begin{equation}
  \zeta_{00}={\tilde\zeta}_{00}\,\frac{g_0 \lambda_s^2}{\hbar}\!\sqrt{\frac{nI_{0}}{\varepsilon_{0}}}\times
\begin{cases}
    \sqrt{\left(\frac2\pi\right)^{\frac32}\frac{V}{\sigma_{a,\rho}^2\,\sigma_{a,z}}} & \!\!\!\!\!\!\text{,\;\;for ({\bf a}),} \\
    1\text{,} & \!\!\!\!\!\!\text{\;\;for ({\bf b}),}
  \end{cases}\!
\end{equation}
with $\lambda_s=(2\pi)/k_s$.

Note that for large Fresnel number cases ($F=10$) in Fig.~\ref{mod_funs}, $\zeta_{lm}$'s decrease slowly with the mode indices  $(lm)$ increasing. This can be easily understood considering the geometric interpretation of $F\propto\sigma^2/\sigma_z$. Large values of the Fresnel number correspond to moderate axial lengths, $\sigma_z$, of the effective interacting region. In this case, many modes fit the illuminated portion of the ensemble of diameter $\sigma$. Inverting this argument, for small $F$ only the most fundamental modes are strongly coupled and contribute to the scattering process. Hence, we observe a steep drop in the singular values for $F=0.1$ in both regimes.

\section{Conclusions}

We have developed a theoretical model of three-dimensional Raman scattering process in atomic vapors, which is applicable when the number of atomic excitations created is low. Because of its simplicity, it allows to immediately separate the system's spatial and temporal degrees of freedom. Moreover, with the decoherence effects included, we have found the time evolution of the numbers of both the Stokes photons and the spin waves produced in the process. Furthermore, by investigating the temporal coherences of the system, we have demonstrated that despite losses the number of collective atomic excitations can be estimated from the time evolution of the light field with a precision that surpasses the projection noise limit. For the cases of ultra-cold and room-temperature atoms, we have shown how to easily utilize our model to evaluate the spatial mode functions of both the Stokes photons and the spin waves. Although the final form of the eigenmodes had to be computed numerically, we have shown that, when either the atomic density within probe is uniform and the pump beam is of a Gaussian shape or vice versa, the calculation can be analytically pursued till the final decomposition by means of working in the basis of the Laguerre-Gauss functions.

Finally, we point out that the above model can be easily generalized and converted to describe other types of Raman scattering. Most importantly, it can be simply modified to represent the inverse read-out process, in which the emitted photons carry the information that was originally stored by the spin waves already present in the atomic ensemble.

\section*{Acknowledgments}

This work was supported by the Foundation for Polish Science TEAM project co-financed by the EU European Regional Development Fund and FP7 FET project Q-ESSENCE (Contract No. 248095). J. Ch. acknowledges the financial support of the National Science Centre. W.W. was also supported by the Polish Government scientific grant (2010-2012).

\appendix

\section{"Time slicing" method}\label{t_slice}

To derive a non-perturbative solution of Eqs.~(\ref{dalm_dt}) and (\ref{dblm_dt}), we first integrate Eq.~(\ref{dblm_dt}) perturbatively in an interval $t\in[0,\tau]$. The result obtained is substituted into Eq.~(\ref{dalm_dt}), which is then integrated in the same time interval. This way, we obtain the output field of scattered photons, denoted by $\hat a^{\rm out}_{lm,1}$, together with $\hat b_{lm,1}$ that are valid at time $\tau$. At the beginning of the next interval $t\in[\tau,2\tau]$, we assume that another unperturbed Stokes field $\hat a^{\rm in}_{lm,2}$ contributes to the process, as if it entered the probe. This operator together with $\hat b_{lm,1}$ defines the new initial conditions for the repetition of the procedure in the next ``time slice'': $t\in[\tau,2\tau]$. Replicating the method iteratively, up to the $j$-th consecutive interval, we obtain the field operators:
\begin{subequations}\label{discr}
  \begin{eqnarray}
    \hat a^{\rm out}_{lm.j}&=&\hat a^{\rm in}_{lm,j-1}+\sqrt{\tau}\,\zeta_{lm}\!\left(1+\frac{\zeta_{lm}^2\tau}{2}\right)^{j-1}\hat b_{lm,0}^\dagger\ \ \ \ \ \ \ \ \ \ \nonumber\\
    &+&\tau\,\zeta_{lm}^2\sum_{i=0}^{j-2}\left(1+\frac{\zeta_{lm}^2\tau}{2}\right)^{j-i-2}\!\!\!\hat a^{\rm in}_{lm,i}\\
    \hat b_{lm,j}&=&\left(1+\frac{\zeta_{lm}^2\tau}{2}\right)^j\hat b_{lm,0}\ \ \ \  \ \nonumber\\
    &+&\sqrt{\tau}\,\zeta_{lm}\sum_{i=0}^{j-1}\left(1+\frac{\zeta_{lm}^2\tau}{2}\right)^{j-i-1}\left[\hat a^{\rm in}_{lm,i}\right]^{\dagger}.
  \end{eqnarray}
\end{subequations}
Effectively, the whole field of scattered photons is divided into slices of length $c\tau$. Although at each step, the solutions are perturbative, we obtain an exponential growth of the effective number of the Stokes photons produced due to the constant accumulation of atomic excitations over time.

Finally, we construct the field operators, which continuously change in time, that is,
\begin{eqnarray*}
  \hat a^{\rm in/out}_{lm}(t)&=&\lim_{j\rightarrow\infty}\left\{\sqrt{\frac jt}\ \hat a^{\rm in|out}_{lm,j}\right\}\\ 
  \hat b_{lm}(t)&=&\lim_{j\rightarrow\infty}\hat b_{lm,j}\,,
\end{eqnarray*}
hence we obtain the set of non-perturbative Eqs. (\ref{fina}) and (\ref{finb}).

\section{Singular value decomposition (SVD)}\label{decomph}

To discretize the description of the system, we seek a functional basis, in which the eigenmodes $\psi_{lm}^{\rm ph}(\vecb r)$ and $\psi_{lm}^{\rm at}(\vecb r)$, introduced in Eq.~(\ref{ft_svd}), could potentially be expressed. Investigating, their definition 
we explicitly perform the Fourier transform of Eq.~(\ref{svd}). Focusing on the photonic $(lm)$ eigenmode, we obtain
\begin{eqnarray}
\psi_{lm}^{{\rm ph}}(\vecb r) & = & \int\!\!d\vecb k\,e^{i\mathbf{k}\cdot\mathbf{r}}\frac{c\,\tau}{\pi k_s^2}\sinc(\frac{\omega_{k}\tau}{2})\,\psi_{lm}^{{\rm ph}}(k_{\varphi},k_{\theta})\nonumber\\
 & \approx & \int\!\!d\vecb k\,e^{i\mathbf{k}\cdot\mathbf{r}}\frac{\delta(k-k_{s})}{2 \pi k^2_s}\sum_{m^{\prime}}\!\psi_{lm;m^{\prime}}^{{\rm ph}}(k_{\theta})\,e^{ik_{\varphi}m^{\prime}}\nonumber \\
 & = &\int^\pi_0\!\!\!\! d k_\theta \sin\! k_\theta \!\!\int^{2\pi}_0\!\!\!\!\!d k_\varphi \,e^{ik_{s}\left[\sin \!k_{\theta}\!\cos(k_{\varphi}-\varphi)\rho+\cos \!k_{\theta}z\right]}\nonumber \\
 & \times & \sum_{m^{\prime}}\psi_{lm;m^{\prime}}^{{\rm ph}}(k_{\theta})\,e^{ik_{\varphi}m^{\prime}},
\end{eqnarray}
where we expanded  $\psi_{lm}^{\rm ph}$ in a multipole series and, as before, assumed that the spectral factor just fixes $k\!=\!|\vecb k|\!\approx\! k_s$. Furthermore, we restrict ourselves to Stokes photons produced in the positive direction at small angles, for which $\psi_{lm;m^{\prime}}^{{\rm ph}}$ is narrowly peaked around $k_\theta=0$. Therefore, we can apply the paraxial approximation, in which $\sin k_\theta\!\approx\!k_\theta$, $\cos k_\theta\!\approx\!1\!-\!\frac{k^2_\theta}{2}$, and extend the polar angle integral's range to $k_\theta\!\in\![0,\infty[$. Hence, with $\alpha\!=\!\varphi\!-\!k_\varphi$,
\begin{eqnarray}\label{psi_ph_par_approx}
\psi_{lm}^{{\rm ph}}(\mathbf{r}) & \approx & e^{ik_{s}z} \sum_{m^{\prime}}e^{i\varphi m^{\prime}}\!\int^\infty_0\!\!\!\! d k_\theta \,k_\theta\, e^{-i\frac{k_{s}k_{\theta}^{2}}{2}z}\,\psi_{lm;m^{\prime}}^{{\rm ph}}(k_{\theta})\nonumber\\
 &\times&\;\int^{2\pi}_0\!\!\!\!\!d\alpha \,e^{ik_{s}k_{\theta}\!\cos\!\alpha\rho}e^{i \alpha m^{\prime}}\nonumber\\
 & = &e^{ik_{s}z} \sum_{m^{\prime}}e^{i\varphi m^{\prime}} \psi_{lm; m^\prime}^{\rm ph} (\rho,z),
\end{eqnarray}
where according to the Hankel transform
\begin{eqnarray}\label{ph_m'}
\psi_{lm;m^\prime}^{{\rm ph}}(\rho,z)&=&2\pi\,i^{m^\prime}\\
&\times&\int^\infty_0\!\!\!\! d k_\theta \,k_\theta \,e^{-i\frac{k_{s}k_{\theta}^{2}}{2}z}\psi_{lm;m^{\prime}}^{{\rm ph}}(k_{\theta})\,J_{m^\prime}(k_s k_\theta \rho)\nonumber
\end{eqnarray} 
and $J_m$ are the Bessel functions of the first kind. 
Similarly, we follow the above procedure in cylindrical coordinates for spin waves. In that case we move to the reference frame used in the derivation of Eq.~(\ref{h_kK_dimless}), in which the eigenmodes are shifted relatively to the ones of Eq.~(\ref{svd}) via $\tilde{\psi}_{lm}^{{\rm at}}\!({\vecb K})\!=\!{\psi}_{lm}^{{\rm at}}\!(\vecb K\!+\!{\vecb k}_p\!-\!k_s\vecb e_z)$.  
As there are no constraints on $\vecb K$, we obtain a general expression for the $(lm)$ eigenmode in position space
\begin{eqnarray}
\psi_{lm}^{\rm at}(\mathbf{r})&=&\!\int\!\!d\vecb K\,e^{i\vecb K\cdot\vecb r}{\psi}_{lm}^{{\rm at}}(\vecb K)\nonumber\\
&=&e^{-i k_s z}\,e^{i{\vecb k}_p\cdot\vecb r}\!\!\int\!\!\!d\vecb K\,e^{i\vecb K\cdot\vecb r}{\tilde\psi}_{lm}^{{\rm at}}(\vecb K)\nonumber\\
&=&e^{-i k_s z}\,e^{i{\vecb k}_p\cdot\vecb r}\sum_{m^{\prime}}\!e^{i\varphi m^{\prime}}{\tilde\psi}_{lm;m^{\prime}}^{\rm at}(\rho,z)
\end{eqnarray}
with
\begin{equation}\label{at_m'}
{\tilde\psi}_{lm;m^{\prime}}^{\rm at}(\rho,z)=2\pi\, i^{m^{\prime}}\!\!\!\int^\infty_0\!\!\!\!\! dK_{\rho}K_{\rho}\,{\tilde\psi}_{lm;m^{\prime}}^{{\rm at}}(K_\rho ,z)\,J_{m^{\prime}}(K_{\rho}\rho)
\end{equation}
and ${\tilde\psi}_{lm;m^{\prime}}^{{\rm at}}(K_\varrho ,z)\!=\!\int^{\infty}_{-\infty}\!dK_{z}\,e^{iK_{z}z}{\tilde\psi}_{lm;m^{\prime}}^{{\rm at}}(K_\varrho ,K_z)$.

In the case of photons, we  fixed $|\vecb k|$ of $\psi_{lm}^{{\rm ph}}(\vecb k)$ mode and applied the paraxial approximation in the position space. Hence, the $z$ dependence of the corresponding cylindrically symmetric $\psi_{lm;m^\prime}^{{\rm ph}}(\rho,z)$ must be exactly the one of the Laguerre-Gauss mode functions, that is,
\begin{equation} \label{ph_z_of_LG}
\underset{l',m'}{\forall}\!:\;\frac{\partial}{\partial z}\left[\int\!\!\rho\, d\rho\left[{\rm LG}_{l'}^{m'}\!(\rho,z)\right]^{\star}\psi_{lm;m'}^{{\rm ph}}(\rho,z)\right]=0.
\end{equation}
We verify the above claim by substituting for the photonic modes of Eq.~(\ref{ph_m'}) into Eq.~(\ref{ph_z_of_LG}) with the Laguerre-Gaussian modes defined as
\begin{eqnarray} \label{LG}
&&\mathrm{LG}^m_{l}\!(\vecb r)=e^{i m \varphi}\,\mathrm{LG}^m_{l}\!(\rho,z)=\frac{e^{i m \varphi}}{w}\sqrt{\frac{2}{\pi}}\sqrt{\frac{l!}{(m+l)!}}\,\,\;\;\;\;\;\;\;\nonumber\\
&&\;\times\,\frac{(1-i\xi_{s})^{l+\frac{m}{2}}}{(1+i\xi_{s})^{l+\frac{m}{2}+1}}\,\rho_s^{m}\exp\!\left[-\frac{\rho^{2}}{w^{2}(1+i\xi_{s})}\right]\text{L}_{l}^{m}\!(\rho_s^{2}),
\end{eqnarray}
where the reduced radial and propagation distances respectively read
\begin{equation*}
 \rho_s=\frac{\sqrt{2}\rho}{w\sqrt{1+\xi_{s}^{2}}}\ \ \ \ \ {\rm and}\ \ \ \ \ \xi_{s}=\frac{2z}{k_sw^{2}}.
\end{equation*} 
The $\mathrm{LG}$ modes form a complete orthonormal basis in the radial plane at any $z$, that is,
\begin{equation*}
\underset{z}{\forall}\!:\;\int \!\! d \boldsymbol \rho\,\left.{\rm LG}^{m}_{l}\!(\vecb r)\!\right.^{\star}\,{\rm LG}^{m'}_{l'}\!(\vecb r)=\delta_{l,l'}\,\delta_{m,m'}.
\end{equation*}
Although, in the case of atoms, neither the magnitude of momentum is fixed nor the paraxial approximation holds, it is also correct to use the Laguerre-Gaussian modes as the basis, but independently for each $z$. 

Finally, we are able to write down a general basis expansion of the eigenmodes of Eq.~(\ref{ft_svd}):
\begin{subequations}\label{modfun_pos}
\begin{eqnarray}
\psi_{lm}^{{\rm ph}}(\mathbf{r})&=&\sum_{l'm'}c^{\rm ph}_{lm;l'm'}\; f^{\rm ph}_{l'm'}\!(\vecb r)\\
\psi_{lm}^{\rm at}(\vecb r)&=&  \sum_{l'm'} c^{\rm at}_{lm;l'm'}(z)\; f^{\rm at}_{l'm'}\!(\vecb r),
\end{eqnarray}
\end{subequations}
with the photonic and atomic basis functions defined as
\begin{equation*}
f^{\rm ph}_{lm}\!(\vecb r)\!=\!e^{ik_{s}z}\,{\rm LG}^{m}_{l}\!(\vecb r)\;\;\;\;\;\;f^{\rm at}_{lm}\!(\vecb r)\!=\!e^{-ik_{s}z}e^{i{\vecb k}_p\cdot\vecb r}\left.{\rm LG}^{m}_{l}\!(\vecb r)\!\!\right.^\star\!\!.
\end{equation*}

Having chosen a functional basis, we decompose the left-hand side of Eq.~(\ref{ft_svd}) and obtain its matrix representation:
\begin{eqnarray}\label{mat_el}
&&\int\!\!d\boldsymbol{\rho}\,d\boldsymbol{\rho}'\!\left.f^{\rm ph}_{lm}(\vecb r)\!\right.^{\star}\mathcal{F}\!\left[\frac{c\,\tau}{\pi k_{s}^{2}}\,\mathrm{sinc}\!\left(\frac{\omega_{k}\tau}{2}\right)h(\mathbf{k},\mathbf{K})\right]\!\!\left.f^{\rm at}_{l'm'}(\vecb r')\!\right.^\star\!\nonumber\\
&&\;\,\approx\!\!\int\!\!d\boldsymbol{\rho}\,d\boldsymbol{\rho}'\!\left.\mathrm{LG}_{l}^{m}\!(\vecb r)\!\right.^{\star}\!e^{-i k_sz}\!\!\int\!\!d\mathbf{k}\,d\mathbf{K}\,e^{i\mathbf{k}\cdot\mathbf{r}}e^{i\mathbf{K}\cdot\mathbf{r}^{\prime}}\frac{\delta(k-k_{s})}{2\pi k_{s}^{2}}\nonumber\\
&&\;\;\;\;\;\;\;\times\; h(\vecb k, \vecb K)\,e^{ik_sz'}\,e^{-i\vecb{k}_{p}\cdot{\vecb r}'}\,\mathrm{LG}_{l'}^{m'}\!(\vecb r'),
\end{eqnarray}
fixing again the Stokes photons momentum at $|\vecb k|\!=\!k_s$ and substituting for the coupling function according to Eq.~(\ref{h_kK_gen}), that is,
\begin{equation}
h(\vecb k, \vecb K)=\frac{1}{(2\pi)^{3}}\!\!\int\!\! d\mathbf{r}\,e^{-i(\mathbf{k}+\mathbf{K}-\vecb{k}_{p})\cdot\mathbf{r}}\;h(\rho,z),
\end{equation}
with $h(\rho,z)=\frac{g_{0}}{\hbar}\,\exp\!\left[-\frac{\rho^{2}}{\sigma^{2}(1+i\xi)}\right]Z(z)$.

The integral (\ref{mat_el}) corresponds to the matrix element $h_{lm;l'm'}$ obtained by projecting onto the $f^{\rm ph}_{lm}$ and $f^{\rm at}_{l'm'}$ basis functions. Since $h(\rho,z)$ is cylindrically symmetric, the matrix elements are found to be diagonal in the mode number $m$, that is,
\begin{equation}\label{hzz'}
h_{lm;l'm'}(z;\!z')=\delta_{m,m'}\,h_{lm;l'm}(z;z')\approx\delta_{m,m'}\,h^m_{l;l'}(z'),
\end{equation}
where
\begin{eqnarray}\label{hz'}
h^m_{l;l'}(z')&=&(2\pi)^{2}\!\int^\infty_0\!\!\!\!\rho\,d\rho\!\int^\infty_0\!\!\!\!\rho'd\rho'\left.\mathrm{LG}_{l}^{m}\!(\rho,z)\!\right.^{\star}\nonumber\\
&\times&\!\!\int^\infty_0\!\!\!\!\!\! k_{\theta}dk_{\theta}\,J_{m}(k_{s}k_{\theta}\rho)\,e^{-i\frac{k_{s}k_{\theta}^{2}}{2}(z-z')}J_{m}(k_{s}k_{\theta}\rho')\nonumber\\
&\times&h(\rho',z')\;\mathrm{LG}_{l'}^{m'}\!(\rho',z').
\end{eqnarray}
In the last step of Eq.~(\ref{hzz'}), similarly to Eq.~(\ref{psi_ph_par_approx}), we employed the paraxial approximation in the case of photons. As proven in Eq.~(\ref{ph_z_of_LG}), the result (\ref{hz'}) is independent of the photonic axial coordinate. Whence, not only $h^m_{l;l'}$ is a function only of $z'$, but also the $z$ coordinate can be arbitrarily fixed. Taking $z\!=\!z'$, we obtain
\begin{equation}\label{h_mll'_final}
h^m_{l;l'}(z')=\lambda_s^{2}\!\int^\infty_0\!\!\!\!\!\!\rho'd\rho'\left.\mathrm{LG}_{l}^{m}\!(\rho',z')\!\right.^{\star}h(\rho',z')\,\mathrm{LG}_{l'}^{m}\!(\rho',z'),
\end{equation}
where $\lambda_s\!=\!2\pi/k_s$ and we have used the identity 
\begin{equation*}
\underset{m}{\forall}:\;\int^\infty_0\!\!\! t\,dt\,J_{m}(t\rho)J_{m}(t\rho')=\frac{\delta(\rho-\rho')}{\rho}.
\end{equation*}
The integral in Eq.~(\ref{h_mll'_final}) can be computed analytically, as an overlap of a single Laguerre-Gaussian mode with a product of a Gaussian and a Laguerre-Gaussian functions \cite{Bandres2010}, that is,
\begin{eqnarray} \label{hz_fin}
h^m_{l;l'}(z)& = & \frac{g_0\lambda^2_s}{\hbar}\,Z(z)\,{(2\chi)}^{m+1}\sqrt{\binom{m+l}{l}\binom{m+l'}{l'}}\nonumber\\
 & \times & \frac{\left(1+i\xi\right)^{1+m}\left(1+\xi_{s}^{2}\right)^{l+l'}}{\left[1+2(1+i\xi)\chi+\xi_{s}^{2}\right]^{1+m+l+l'}}\left(\frac{1-i\xi_{s}}{1+i\xi_{s}}\right)^{l'-l}\nonumber\\
 & \times & \,_{2}\text{F}_{1}\!\left[-l,-l';m+1;4\chi^{2}\left(\frac{1+i\xi}{1+\xi_{s}^{2}}\right)^{2}\right]\!,
\end{eqnarray}
where $\chi\!=\!\sigma^{2}/w^{2}$ and $_2\text{F}_1$ is the Gaussian hypergeometric function. This result resembles the one calculated for the eigenmode problem in spontaneous parametric down-conversion \cite{Miatto2011}.  Furthermore, the $z$ dependence can also be discretized in the experimentally valid region, so that the overall matrix defined in Eq.~(\ref{hz_fin}) fully describes the system in a finite dimensional form. For example, in the cases discussed in the paper, by fixing the width parameter of the LG modes to $w\!=\!\sqrt{\sigma_{a,z}/k_s}$ in the ultra-cold regime ({\bf a}), and $w\!=\!\sqrt{L/k_s}$ in the room-temperature regime ({\bf b}), we can reparametrize the problem and, rather than the relevant ranges $|z|\!<\!3\sigma_{a,z}$ and $|z|\!<\!\frac{L}{2}$, discretize the propagation distances, $\xi_s$, in $|\xi_s|\!<\!6$ and $|\xi_s|\!<\!1$ respectively.

Finally, the continuous integral eigenmode problem of Eq.~(\ref{svd}) is reduced to performing the singular value decomposition (SVD) of the matrix:
\begin{equation}
h^m_{l;l'}(z)=\sum_{k} \zeta_{k m}\,c^{\rm ph}_{k m;l m}\, c^{\rm at}_{k m; l' m}\!(z),
\end{equation}
where the diagonality of $h_{lm;l'm}$ in $m$ (the cylindrical symmetry of the coupling function) fixes completely the azimuthal angle's mode number. Hence, the expressions for photons and spin waves eigenmodes of Eq.~(\ref{modfun_pos}) simplify to the form of Eq.~(\ref{mod_pos_exp}) stated in the paper
\begin{eqnarray*}
    \psi^{\rm ph}_{lm}(\vecb r)&=&e^{ik_sz}\,e^{im\varphi}\,\psi^{\rm ph}_{lm}(\rho,z)\\
    \psi^{\rm at}_{lm}(\vecb r)&=&e^{-ik_sz}\,e^{i{\vecb k}_p\cdot\vecb r}\,e^{-im\varphi}\,\psi^{\rm at}_{lm}(\rho,z).
\end{eqnarray*}
However, now, having computed the coefficients $c^{\rm ph/at}_{lm;l'm}$, we can reconstruct the effective eigenmodes as
\begin{subequations}
  \begin{eqnarray}
    \psi^{\rm ph}_{lm}(\rho,z)&=&\sum_{l'}c^{\rm ph}_{lm;l'm}\;{\rm LG}^{m}_{l'}\!(\rho,z)\\
    \psi^{\rm at}_{lm}(\rho,z)&=&\sum_{l'} c^{\rm at}_{lm;l'm}(z)\, \left.{\rm LG}^{m}_{l'}\!(\rho,z)\!\right.^\star\!\!.
  \end{eqnarray}
\end{subequations}

\bibliographystyle{apsrev4-1}
\bibliography{raman2}

\end{document}